\newcounter{myctr}

\documentclass{ws-ijqi}

\begin{document}

\markboth{H.-T. Elze}
{Ontological states and dynamics of discrete (pre-)\,quantum systems}

\catchline{}{}{}{}{}

\title{ONTOLOGICAL STATES AND DYNAMICS OF DISCRETE (PRE-)\,QUANTUM SYSTEMS  
}

\author{HANS-THOMAS ELZE 
}

\address{Dipartimento di Fisica ``Enrico Fermi'', Universit\`a di Pisa, \\ 
Largo Pontecorvo 3, I-56127 Pisa, Italia
\\ elze@df.unipi.it}



\maketitle


\begin{abstract}
The notion of ontological states is introduced here with reference to the  
{\it Cellular Automaton Interpretation of Quantum Mechanics} proposed by G. 't\,Hooft. 
A class of discrete deterministic ``Hamiltonian'' Cellular Automata is defined that has been shown to bear many features in common with continuum quantum mechanical models, however, deformed by the presence of a finite discreteness scale $l$, such that for $l\rightarrow 0$ the usual properties result -- {\it e.g.}, concerning linearity, dispersion relations, multipartite systems, and Superposition Principle.  
We argue that within this class of models only very primitive realizations of ontological states and their dynamics can exist, since the equations of motion tend to produce superposition states that are not ontological. The most interesting, if not only way out seems to involve interacting multipartite systems composed of two-state ``Ising spins'', which evolve by a unitary transfer matrix. Thus, quantum like and ontological models appear side by side here, but distinguished by second-order and first-order dynamics, respectively. 
\end{abstract}




\section{Introduction}

Debates concerning the interpretation(s) of quantum mechanics and the cornerstones of its conceptual foundations have been around since times when an apparently final formulation was reached, almost a century ago.\,\cite{review1} Only with the rapid advances of experimental capabilities in recent decades, aiming for quantum metrology and quantum information processing in particular, related questions have come to the forefront 
(again).\,\cite{review2} ``Reconstructions'' of quantum theory from various alternative sets of axioms, without changing its contents, are shedding light on the alleged ``quantum weirdness'' by offering new perspectives for experiments.   

Nevertheless, problems related to the very foundations of quantum mechanics persist, such as the measurement problem, the nature of the seemingly intrinsic randomness accompanying otherwise deterministic unitary quantum processes, the unresolved incompatibility with General Relativity, the coupling between truly classical and quantum systems (see, {\it e.g.} Ref.\,\cite{hybrids} and references therein), {\it etc.} This has led to attempts to go beyond quantum theory by ``deforming'' in one way or another its structural elements. The present work continues research in this direction, especially by looking for ontological states in simple pre-quantum systems. 

Ontological states that underlie quantum and, {\it a fortiori}, classical states of physical objects have been 
discussed as a working hypothesis of the {\it Cellular Automaton Interpretation} of Quantum Mechanics (QM) proposed by G. 't\,Hooft.\,\cite{tHooft2014} 

We recall the motivation to re-examine the foundations of quantum theory in perspective of essentially classical concepts -- above all, determinism and existence of ontological states of reality -- which stems from observations of quantum features in a large variety of deterministic and, in some sense, ``classical'' models. The Born rule and collapse of quantum  states in measurement processes find a surprising and intuitive explanation here, where quantum states are mathematical objects representing fictitious superpositions of ontological (micro) states, while classical states are ontological (macro) states, or probabilistic superpositions thereof, respectively refering to vastly different scales in nature.\,\cite{tHooft2014} 

It is worth emphasizing that quantum states here are considered to form part of the mathematical language used, they are ``templates'' for the description of the ``reality beneath'', including ontological states and their deterministic dynamics. 

Most of related pre-quantum models considered, so far -- see, for example,   
Refs.\,\cite{H1,H2,H3,Kleinert,Elze,Groessing,Khrennikov,Margolus,Jizba,Mairi,Isidro,DAriano,Wetterich16} 
and  references therein -- cannot be 
generalized in a straightforward way to cover real phenomena, say, as in the Standard Model, incorporating interactions and relativity. 

However, finite and discrete Cellular Automata (CA), as invoked by the  {\it Cellular Automaton Interpretation} (CAI), may provide the necessary versatility.\,\cite{tHooft2014} We shall presently continue to study the class of Hamiltonian CA.\,\cite{PRA2014,EmQM13} Our aim is to demonstrate that even these apparently simplest CA, in particular multipartite ones,\,\cite{IJQI16} allow for ontological states and their evolution, besides (proper) quantum ``template'' states (especially in the continuum limit). 

To embark on this, we represent here some essential aspects of ontological states, according to CAI,\,\cite{tHooft2014} which serves to obtain an operational understanding of what we should be looking for, in the following:   
\vskip 0.2cm\noindent 
{\it ONTOLOGICAL STATES} ($\cal OS$) are states a deterministic physical system can be in. They are denoted by    
$|A\rangle ,\; |B\rangle ,\; |C\rangle ,\;\dots\;$. The set of all states may be very large, but is assumed to be denumerable, for simplicity.
\vskip 0.1cm\noindent 
There exist {\it no superpositions} of $\cal OS$ ``out there'' as part of physical reality. 
\vskip 0.1cm\noindent  
The $\cal OS$ evolve by {\it permutations} among themselves,  
$\dots\rightarrow |A\rangle\rightarrow |D\rangle\rightarrow |B\rangle\rightarrow\;\dots\;,$ for example. Apparently this is the only possibility, besides  producing a growing set of states or superpositions, which do not belong to the initial set of $\cal OS$.  
\vskip 0.1cm\noindent 
By declaring the $\cal OS$ to form an orthonormal set, fixed once for all, a {\it Hilbert space} can be defined.  -- Operators which are diagonal on this set of 
$\cal OS$ are {\it beables}. Their eigenvalues describe physical properties of the $\cal OS$, corresponding to the 
abstract labels $A,\; B,\; C,\;\dots\;$ above.   
\vskip 0.2cm\noindent  
{\it QUANTUM STATES} ($\cal QS$) are superpositions of $\cal OS$. These are   
templates for doing  physics with the help of mathematics. --  
The amplitudes specifying superpositions need to be interpreted, when applying the formalism to describe experiments. 
Here the {\it Born rule is built in},  {\it i.e.}, by definition! By experience, interpreting amplitudes in terms of probabilities has been an amazingly useful invention.\footnote{\,'t\,Hooft has convincingly argued for this point of view, considering also that the proportionality  between {\it absolute values squared} of complex amplitudes and probabilities does not have to be this way, which is singled out by mathematical convenience.\,\cite{tHooft2014}}   
\vskip 0.2cm
From here the machinery of QM departs smoothly, including especially the powerful techniques related to unitary transformations. The latter exist, in general, only in discrete form on the level of $\cal OS$, due to the absence of superpositions. 

However, we must be aware of  roadblocks, when trying to built a theory based on CAI and  $\cal OS$, in particular.   
  
Namely, it has turned out to be very difficult to find an ontological model for any part of the Standard Model, other than for free particles. --    
In fact, it was unknown for a long time how to incorporate interactions into any kind of ontological toy model.    

Furthermore, it is generally not obvious how to connect evolution of $\cal OS$ by permutations to a Hamiltonian operator. This is desirable in view of the unitary evolution of 
$\cal QS$, which is  to be recovered.  

Concluding this brief introduction of some essential points of CAI, one more remark is in order, concerning the absence of the infamous ``measurement problem''.\footnote{\,For a particularly clear criticism of earlier attempts to solve the measurement problem of and within standard QM, {i.e.} without modifying its basic tenets (such as in stochastic modifications of the Schr\"odinger equation), see, {\it e.g.},  
Ref.\,\cite{Adler03}} This provides additional motivation for pursuing the consequences of an approach based on ontological states:  
\vskip 0.2cm\noindent  
{\it CLASSICAL STATES} of a macroscopic deterministic system, including billiard balls, pointers of apparatus, planets, are {\it probabilistic distributions} of $\cal OS$, since  any kind of repeated experiments performed by physicists, with only limited control of the circumstances, pick up different initial conditions regarding the $\cal OS$. Hence, different outcomes of apparatus readings must generally be expected. Yet any reduction or collapse to a $\delta$-peaked distribution, say, of pointer positions is only an apparent effect, induced by the intermediary use of quantum mechanical  templates in describing the evolution of $\cal OS$ during an experiment. Ontologically speaking, there were / are no superpositions, to begin with, which could possibly collapse!\,\cite{tHooft2014}
\vskip 0.2cm 
Indeed, quantum theory has been very successful in describing experiments. Its {\it linearity} is the characteristic feature of the unitary dynamics embodied in the Schr\"odinger equation. For a prospective ontological theory, it is of interest that QM  remains notoriously  indifferent to any reduction or collapse process one might be tempted to add on, in order to modify the collapse-free linear evolution. --  This linearity does not depend on the particular object under study, provided it is sufficiently isolated from 
anything else. It is expressed in the Superposition Principle and   
entails interference effects and the possibility of nonclassical correlations among parts of   
composite objects, {\it i.e.} entanglement in multipartite systems.    

Considering a {\it discrete} dynamical theory that apparently deviates drastically from quantum theory, we have recently shown,   
with the help of Sampling Theory,\,\cite{Shannon,Jerri} how   
members of this class of {\it Hamiltonian CA} can be mapped one-to-one to continuum models belonging to nonrelativistic QM, which   
are characteristically deformed by the presence of a {\it fundamental length or time scale}.\,\cite{PRA2014,EmQM13}  
This mapping is only consistent, if   
the action for the variational principle of the discrete dynamics is harmonic, on one side,  
and the quantum mechanical continuum description is local in time, on the other, implying that both theories {\it have to be linear}.\,\cite{Discrete14}  

In the following Section\,2., we summarize the quantum features of Hamiltonian CA, in order to make this article selfcontained and introduce some results to which we shall refer  later. In Section\,3., we consider the formal solution of the CA equations of motion, in analogy to solving the Schr\"odinger equation by exponentiation of the Hamilton operator. Thus, we can address the question whether there is any room for $\cal OS$. A simplistic example shows that this is indeed the case. In order to allow for more interesting dynamics, we study interacting multipartite systems in Section\,4. In the concluding section, we discuss our findings.    

\section{The quantumness of Hamiltonian CA}  

Early on T.\,D. Lee, among others, suggested that it should be worthwhile to reconsider the foundations of (quantum) physics in the light of the hypothesis that we have to deal with a fundamental discreteness of nature:\,\cite{TDLee}    
\vskip 0.2cm\noindent 
There is a fundamental {\it time or length scale} $l$ ($c=1$)  
``such that in a ($d$+1)-dimensional spacetime volume $\Omega$ maximally   
$\Omega/l^{d+1}$ measurements can be performed or maximally   
this number of events can take place''.  -- 
Thus, {\it dynamics is discrete} and time may be dynamical.  
\vskip 0.2cm
Following this suggestion, we have considered discrete mechanical systems, in particular  discrete ``clocks'' interacting with ordinary QM ones.\,\cite{DMTimeMach12} Which has quickly led to study consequences of such discreteness for QM objects themselves or, rather, the deformation of their quantum mechanical properties by the presence of the  discreteness scale. This has been the beginning of {\it Hamiltonian CA}, which describe simple discrete dynamical  systems that show quantum features --  especially, but not only in the continuum limit.\,\cite{PRA2014,EmQM13} 

These CA are ``bit processors'', {\it i.e.} classical CA with  denumerable  degrees of freedom. Their state is described by {\it integer valued} coordinates $x_n^\alpha$ 
and momenta $p_n^\alpha$, where  
$\alpha\in {\mathbf N_0}$ labels different degrees of freedom and  
$n\in {\mathbf Z}$ successive states.  

Only {\it finite differences} of variables, such as $\Delta f_n:=f_n-f_{n-1}$,  
are to play a role here, where {\it no infinitesimals} or ordinary derivatives are available!

The dynamics and symmetry properties of such systems are conveniently encoded in an {\it action principle}.\,\cite{PRA2014,EmQM13} However, the ensuing CA updating rules or equations of motion are only consistent, if the form of admissible actions is suitably 
constrained.\,\cite{Discrete14,Wigner13} 
Ultimately, this is responsible for the {\it linearity} of QM models, which are obtained in the continuum limit within this class of CA.  

The action principle yields  the   
{\it finite difference} equations of motion: 
\begin{eqnarray}\label{eomx} 
\dot x_n^\alpha &=& S_{\alpha\beta}p_n^\beta +A_{\alpha\beta}x_n^\beta  
\;\;, \\ [1ex] \label{eomp}
\dot p_n^\alpha &=& -S_{\alpha\beta}x_n^\beta +A_{\alpha\beta}p_n^\beta  
\;\;, \end{eqnarray}  
where the integer-valued symmetric and antisymmetric matrices,
$\hat S\equiv \{S_{\alpha\beta}\}$  and $\hat A\equiv\{A_{\alpha\beta}\}$, respectively, 
define a particular model. Here we introduced the notation 
$\dot O_n:=O_{n+1}-O_{n-1}$, in order to guide the eye and to indicate the analogy with Hamilton's equations in the continuum; hence the name {\it Hamiltonian  CA}.   

The Eqs.\,(\ref{eomx})-(\ref{eomp}) are invariant under reversal of the updating direction (discrete time reversal invariance), since they allow to 
update the state of a CA in both directions, 
$(n\mp 1,n)\;\rightarrow\;(n\pm 1)$, given appropriate initial values of the variables.   

Remarkably, both equations of motion can be combined into one: 
\begin{equation}\label{discrSchroed}
\dot\psi_n^\alpha\;=\;-iH_{\alpha\beta}\psi_n^\beta
\;\;, \end{equation} 
plus its adjoint, by introducing a self-adjoint ``Hamiltonian'' matrix, 
 $\hat H:=\hat S+i\hat A$, and complex integer-valued (also known as Gaussian 
integer) state variables, 
$\psi_n^\alpha :=x_n^\alpha +ip_n^\alpha$. Which obviously resembles the 
Schr\"odinger equation, despite involving only Gaussian integer quantities.   

The resemblance between Eq.\,(\ref{discrSchroed}) and the Schr\"odinger equation in the continuum is no accident. Namely, there is an invertible map between this discrete equation describing Hamiltonian CA in terms of the variables $\psi_n^\alpha$ and a continuous time equation describing the {\it same} CA in terms of a complex ``wave function''  $\psi^\alpha (t)$.\,\cite{PRA2014,EmQM13}  This has been constructed  by applying Sampling Theory,\,\cite{Shannon,Jerri} which introduces the 
fundamental discreteness scale $l$ in terms of a bandwidth limit or high-frequency cut-off  for wave functions.  
\vskip 0.2cm 
In this way, we have obtained the Schr\"odinger equation of QM, however, modified by additional $l$-dependent terms which involve higher-order derivatives with respect to 
time.\,\cite{PRA2014,EmQM13} 

This implies an $l$-dependent {\it dispersion relation} for stationary  states, which may lead to phenomenological consequences for an  
underlying CA model.\,\cite{PRA2014,Wigner13}  Furthermore, there are 
 $l$-dependent {\it conservation laws} in one-to-one correspondence with those of the related  quantum mechanical model, which is obtained in the continuum limit, 
$l\rightarrow 0$.\,\cite{PRA2014} Again, this may have phenomenological consequences in the real world. See Ref.\,\cite{Torino16} for a recent discussion of the discrete and continuum versions of the CA conservation laws.  

If ordinary space is discrete as well, then a {\it Generalized Uncertainty Principle} can be derived from the CA description, based on Robertson's inequality:\,\cite{Robertson29}   
$\Delta \hat A\Delta \hat B\geq |\langle [\hat A,\hat B]\rangle |/2$, with $\hat A$ and $\hat B$ Hermitean operators and where 
$\Delta \hat A^2:=\langle \hat A^2\rangle -\langle \hat A\rangle ^2$, as usual. Here,   
we can define position and momentum operators, respectively, by $X_{mn}:= lm\delta_{mn}$ and   
$P_{mn}:=-i(\delta_{m,n-1}-\delta_{m,n+1})/2l$. This yields:  
$$\Delta \hat X\Delta \hat P\geq\frac{1}{2}|1+\frac{l^2}{2}\langle \hat P^2\rangle |\;\;.$$  
Furthermore, for $\langle P\rangle=0$, this implies a minimal uncertainty 
$\Delta X_{min}=l/\sqrt 2$.\,\cite{DGigli} Due to discreteness, such an effect was to be expected. -- Similar results have been advertised in the context of various quantum gravity models and the resulting phenomenology has recently found considerable attention; see, for example, Ref.\,\cite{GUP} and references therein. 

For our purposes, in the following, another quantum feature will be essential. We have shown that within the class of CA under consideration, one can consistently define {\it multipartite systems}. They obey the Superposition Principle, including the possibility of entanglement, which derives from the tensorial structure of a multipartite Hilbert space in QM.\,\cite{IJQI16} We will make use of this when looking for $\cal OS$, beyond the limitation posed by single systems of few degrees of freedom. 

To conclude this summary of quantum mechanical aspects of Hamiltonian CA, we emphasize that the corresponding standard results of QM are reproduced in the continuum limit, $l\rightarrow 0$, of the {\it deformation} of QM implied here. In retrospect, one could see our results, so far, as pertaining to a particular {\it discretization} of QM, which introduces the scale $l$. However, in the following, we will investigate  
the possible existence of  $\cal OS$ in the present class of discrete models, thus leaving standard QM, as suggested by the {\it Cellular Automaton Interpretation} (CAI).\,\cite{tHooft2014}   

\section{Towards ontological states} 

We recall from the Introduction -- where the concept of ontological states was explained with reference to CAI -- that  $\cal OS$ evolve by permutations among themselves. This is  quite different from the behaviour commonly found in QM, namely evolution by the dynamical formation of superposition states (except for stationary states). 

In order to study the possibility of $\cal OS$ of Hamiltonian CA, we remind ourselves of the general solution of the Schr\"odinger equation, 
$$\partial_t\psi(t)=-i\hat H\psi(t)\;\;\;\Rightarrow\;\;\;\psi(t)=\mbox{e}^{-i\hat Ht}\psi (0)\;\;,$$ 
given the initial state $\psi (0)$, and begin by considering an analogous formal solution of the CA equation of motion (\ref{discrSchroed}),  
$\dot\psi_n=\psi_{n+1}-\psi_{n-1}=-i\hat H\psi_n$. 
Introducing an auxiliary operator $\hat\phi$ through $2\sin\hat\phi :=\hat H$, we find indeed by elementary means: 
\begin{equation}\label{discrsol}
\psi_n=\frac{1}{2\cos\hat\phi}\big (\mbox{e}^{-in\hat\phi}[\mbox{e}^{i\hat\phi}\psi_0+\psi_1]
+(-1)^n\mbox{e}^{in\hat\phi}[\mbox{e}^{-i\hat\phi}\psi_0-\psi_1]\big )
\;\;. \end{equation}  
Here, in general, two initial states have to be given, $\psi_0$ and $\psi_1$, reflecting the fact that  
the Hamiltonian CA are described by a {second-order finite difference equation}.  

With the help of the general solution (\ref{discrsol}), one verifies the following relation: 
\begin{equation}\label{Tsol1}
\psi_n=\hat T(n-m+1)\psi_{m+1}+\hat T(n-m)\psi_{m}  
\;\;, \end{equation} 
where $\hat T$ is a transfer operator that can be read off by comparing with the explicit form 
of this relation; in any case, this generalizes the composition law for the unitary time 
evolution operator in QM. -- Furthermore, we find that the simple exponential form of the solutions in QM can be  
recovered from Eq.\,(\ref{discrsol}) by taking the appropriate limits    
$n\rightarrow\infty$ and $l\rightarrow 0$, keeping $n\cdot l$ fixed,  {\it and} choosing initial conditions such that 
$\psi_1\equiv\psi_0$. In this case, we have:    
\begin{equation}\label{Tsol2} 
\psi_n=\big [\hat T(n+1)+\hat T(n)\big ]\psi_0
\;\;. \end{equation}     
The Eq.\,(\ref{Tsol1}) and especially Eq.\,(\ref{Tsol2}) tell us to expect that superposition states likely will be formed during CA evolution and, therefore, {\it not} ontological states which evolve by permutations among themselves.  

This seems to obstruct the search for $\cal OS$ from the outset.  

\subsection{Can superposition states be avoided? } 

By way of a very simple example, we illustrate that notwithstanding the previous remarks, 
it is possible to have $\cal OS$ in a  {\it two-state} CA. 

Consider the CA described by $\psi_n^\alpha ,\;\alpha =1,2\;$,  with equation of motion given by:  
\begin{equation}\label{exEoM}
\psi_n=\psi_{n-2}-i\hat H_2\psi_{n-1}\;,\;\;
\psi_n\equiv \left (\begin{array}{c}\psi_n^1 \\ \psi_n^2\end{array}\right )\;,\;\; 
\hat H_2:=\left (\begin{array}{c  c} 0 & 1 \\ 1 & 0 \end{array}\right )\equiv\hat\sigma_1
\;\;. \end{equation}   
Furthermore, we choose two orthogonal initial states, 
$\psi_0=(1,0)^t\;\neq\;\psi_1=(0,1)^t$. Then, by simply solving the equation of motion 
iteratively (or by using the general result above), we obtain the following sequence of states:  
\begin{eqnarray}\label{twoditer}
&\;&\psi_0\;,\;\psi_1\;,\;\psi_2=(1-i)\psi_0\;,\;\psi_3=-i\psi_1\;,
\nonumber \\ [1ex] 
&\;&\psi_4=-i\psi_0\;,\;\psi_5=-(1+i)\psi_1\;,\;
\psi_6=-\psi_0\;,\;\psi_7=-\psi_1\;,\;\dots 
\;\;, \end{eqnarray} 
which after four more steps obviously begins to reproduce the initial pair of states. -- 
We note that the normalization of the states, considered as Hilbert space vectors for a moment, 
changes during the evolution. For QM, violating the conservation of the norm of states would be a disaster.  
However, for Hamiltonian CA the norm is not conserved, but is replaced by the  {\it conservation of a two-time  
correlation function}, which appropriately reproduces the norm conservation in the continuum limit.\,\cite{PRA2014,Torino16}

Thus, apart from a presently irrelevant change of normalization, the dynamics produces an ongoing permutation of the 
two orthogonal input states. This is an elementary example where a CA evolves two 
$\cal OS$, as required by the {\it Cellular Automaton Interpretation}.\,\cite{tHooft2014}   
 
\subsection{More complex CA with permutation-like evolution of $\cal OS$?} 

The previous rather primitive example naturally poses the question, whether it can be 
generalized somehow to describe more interesting dynamics.  

Obviously, we might consider systems with a block diagonal $\hat H$ that describes {\it multiple two-state components}.  
However, this yields nothing really new in comparison to the previous example, since the two-state components 
remain indifferent to each other. 

Another option is to enlarge the state space into higher dimensions. In fact, it can be easily worked out that the Hamiltonians: 
\begin{equation}\label{H3H4}
\hat H_3:=\left (\begin{array}{c  c c} 0 & -i & 1 \\ i & 0 & -i \\ 1 & i & 0 \end{array}\right ) \;\;,\;\;\;
\hat H_4:=\left (\begin{array}{c  c c c} 0 & -i & 0 & 1 \\ i & 0 & -i & 0 \\ 0 & i & 0 & -i \\ 1 & 0 & i & 0 \end{array}\right )
\;\;, \end{equation} 
for three- and four-state CA, respectively, lead to analogous evolution-by-permutation of $\cal OS$ as the previous example, $\hat H_2$ of Eq.\,(\ref{exEoM}). 
Interestingly, $\hat H_3$ and $\hat H_4$ do not change the normalization of the states, but introduce only phases when permuting states. Similar results can be 
obtained for state spaces of any finite dimension, it appears, and we will report related aspects elsewhere. 

Instead, we presently turn to the study of {\it interacting multipartite Hamiltonian CA}, which are composed of two-state subsystems. This will be described in the next Section\,4.       

\section{Dynamics = Permutations in multipartite CA}   
 
Completing the exploration of the quantum mechanical features of Hamiltonian CA, 
we have shown that also multipartite CA can be consistently 
formulated.\,\cite{IJQI16,Torino16}. Which allows for the Superposition Principle and the notion of entanglement already at the level of discrete CA models.
 
We recall that two main obstacles had to be overcome, in order to achieve this: 
First, the {\it Leibniz rule} is violated by the second-order finite difference operator featuring prominently in CA equations of motion, such as Eq.\,(\ref{discrSchroed}). We have:  
\begin{equation}\label{Leibniz} 
\dot {[\phi^{(1)}_n\phi^{(2)}_n]}=
\dot \phi^{(1)}_n\textstyle{\frac{\phi^{(2)}_{n+1}+\phi^{(2)}_{n-1}}{2}}+
\textstyle{\frac{\phi^{(1)}_{n+1}+\phi^{(1)}_{n-1}}{2}}\dot \phi^{(2)}_n 
\neq\dot\phi^{(1)}_n\phi^{(2)}_n +\phi^{(1)}_n\dot \phi^{(2)}_n
\;\:, \end{equation}  
applied here to the product of two wave functions.  
Second, the map between discrete CA and corresponding deformed quantum mechanical continuum models, constructed by Sampling Theory, 
when applied directly to products of wave functions, such as $\phi_n^{(1)}\phi_n^{(2)}$, results in nonlocal (in time) expressions of the corresponding $\phi^{(1)}(t)$ and $\phi^{(2)}(t)$. -- 
Both of these effects can produce {\it unphysical correlations} among noninteracting parts of composite systems.\,\cite{IJQI16,Discrete14}  

These problems have been resolved by introducing a 
``{\it many-time formulation}'' -- known as a crucial ingredient of the Tomonaga-Schwinger formulation of relativistic quantum field theory.\,\cite{Dirac,Tomonaga,Schwinger} Applied to the case at hand, this means that each subsystem $i$ has its own proper counter $n_i$ of successive states. Then,  
the finite difference operator and, similarly, the continuum mapping act on these discrete arguments independently. To give an example, the equation of motion describing a bipartite CA can then be stated as follows:  
\begin{equation}\label{biEoM} 
\dot\psi^{\alpha_1\alpha_2}_{\mathbf{n_1}n_2}
+\dot\psi^{\alpha_1\alpha_2}_{n_1\mathbf{n_2}}= 
-iH^{\alpha_1\alpha_2\beta_1\beta_2}\psi^{\beta_1\beta_2}_{n_1n_2}
\;\;, \end{equation}   
where $\dot\psi^{\alpha_1\alpha_2}_{\mathbf{n_1}n_2}:=\psi^{\alpha_1\alpha_2}_{n_1+1\; n_2}-\psi^{\alpha_1\alpha_2}_{n_1-1\; n_2}$ and analogously for the 
second term on the left-hand side. This equation avoids the problem of unphysical correlations.  -- The generalization of this equation for a multipartite system can immediately be written down as well.  
For simplicity of notation, we will stick mostly with the bipartite case in the following. 

\subsection{On the existence and uniqueness of solutions}

We will now take a closer look at the bipartite CA equation of motion (\ref{biEoM}), keeping the multipartite case in mind.   

By resorting to a many-time formulation of the evolution equation, the number of {\it initial conditions} needed to select a particular sequence of CA updates has increased, as compared to a single-time equation. Furthermore, because of the sum of terms appearing on the left-hand side of Eq.\,(\ref{biEoM}), or its multipartite generalization, with finite difference operations acting separately on the counter variables $n_i$, the interpretation of the equation as an updating rule is {\it not unique}, it depends on how initial data are specified and how the updating is performed. 

To illustrate this, consider the two-dimensional lattice of points labelled by all  integer valued pairs $(n_1,n_2)$. If we provide initial values of the wave function 
$\psi^{\beta_1\beta_2}_{n_1n_2}$ on two neighbouring horizontal (vertical) lines of lattice points, the Eq.\,(\ref{biEoM}) can be used to determine the wave function on a third neighbouring horizontal (vertical) line of such points, and so on. Here, a whole line is updated simultaneously, if no additional rule is given. -- One can also propagate the wave function from intial data given on two neighbouring diagonal lines of lattice points plus its value on one extra neighbouring point; in this case, the propagation proceeds from the extra point one-dimensionally along the third neighbouring diagonal defined by this point, 
in both directions unless restricted somehow.  

In these examples, the propagation will, in general, lead to superposition states again and, thus, not be relevant for the search of $\cal OS$. 

This suggests to ammend the equation by imposing a {\it synchronizing constraint}, in order to uniquely specify the behaviour of a multi-time CA for a constrained set of initial conditions.  

The simplest possible constraint could be to set the subsystem state counters equal to each other, $n_1=n_2=\dots$ and add up the resulting identical contributions on the left-hand side of a multipartite CA equation of motion.\,\cite{IJQI16} -- Another possibility is:   
\begin{equation}\label{constraint1} 
\dot\psi^{\alpha_1\alpha_2}_{\mathbf{n_1}n_2}+\dot\psi^{\alpha_1\alpha_2}_{n_1\mathbf{n_2}}= 
\psi^{\alpha_1\alpha_2}_{n_1+1\; n_2+1}-\psi^{\alpha_1\alpha_2}_{n_1-1\; n_2-1}
\;\;, \end{equation}  
for a bipartite system. The generalization for the multipartite case is obvious here and in the following example. --   
Finally, we may implement a synchronization by: 
\begin{equation}\label{constraint2}  
\dot\psi^{\alpha_1\alpha_2}_{\mathbf{n_1}n_2}+\dot\psi^{\alpha_1\alpha_2}_{n_1\mathbf{n_2}}= 
\psi^{\alpha_1\alpha_2}_{n_1+1\; n_2+1}
\;\;, \end{equation}  
which will be most interesting for our purposes. 

\begin{figure} 
\includegraphics[width=1.0\columnwidth, trim=-8mm 0mm -8mm 0mm]{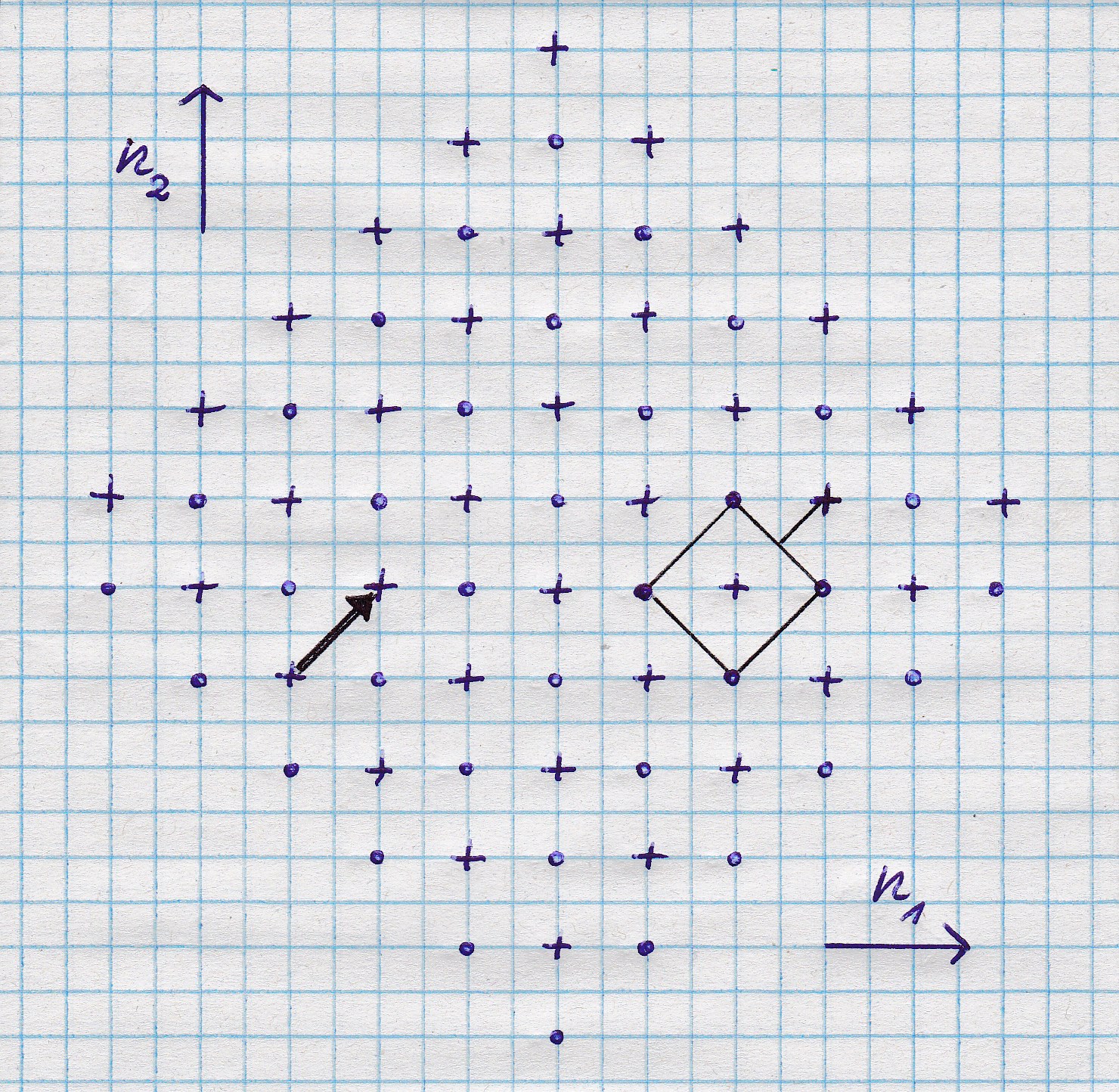}  
\caption{The lattice of two-time points $\{ (n_1,n_2)|n_{1,2}\in\mathbf{Z}\}$, divided into  ``even'' and ``odd'' sublattices marked by $\bullet$ and {\bf +}, respectively. An example is shown of one odd and four even lattice points connected by lines symbolizing the constraint Eq.\,(\ref{constraint2}). The diagonally pointing arrow symbolizes an updating according to Eq.(\ref{firstorder0}).} 
\end{figure} 

Consider the lattice sketched in Figure\,1. It is divided into the ``even'' and ``odd'' sublattices of two-time points. Let the even lattice be sychronized 
according to Eq.\,(\ref{constraint2}). This can be achieved everywhere in such a way that the points corresponding to the left-hand side of this equation form a square on the even lattice, while the point corresponding to the right-hand side lies on the odd lattice, as illustrated. 

With this synchronization, Eq.\,(\ref{constraint2}), the general bipartite CA updating equation (\ref{biEoM}) becomes  
a {\it first-order equation} relating two successive values of the wave function on the 
odd sublattice.  We obtain:   
\begin{equation}\label{firstorder0}
\psi^{\alpha_1\alpha_2}_{n_1+1\;n_2+1}=
-iH^{\alpha_1\alpha_2\beta_1\beta_2}\psi^{\beta_1\beta_2}_{n_1n_2}
\;\;, \end{equation}  
which can be rewritten as:  
\begin{equation}\label{firstorder1}
\psi^{\alpha_1\alpha_2}_{n+1}=
-iH^{\alpha_1\alpha_2\beta_1\beta_2}\psi^{\beta_1\beta_2}_n
\;\;, \end{equation}
since effectively only one variable discrete ``time'' argument $n$ remains, setting $n_1\equiv n+m_1,\;n_2\equiv n+m_2$, with $m_{1,2}$ integer parameters. 

From here on the two sublattices may be forgotten, resulting in a one-time  propagation.  
The existence and uniqueness of {\it solutions} of this synchronized equation follows simply from the fact that the equation can be solved by iteration. 

However, several remarks are in order here. -- The synchronization by such linear constraints, in one way or another, has an effect on the multi-time formulation. It can even lead to a {\it single-time equation}, such as Eq.\,(\ref{firstorder1}), with an obvious generalization for the multipartite case. 

We recall that the multi-time formulation has been necessary, in order to set up equations for multipartite systems which produce quantum mechanical many-body equations in the continuum limit.\,\cite{IJQI16} -- 
Therefore,  
it must not be totally surprising that when looking for $\cal OS$ and their dynamics, we need discrete equations that somehow {\it differ} from those which yield quantum mechanical models. 

The Eq.\,(\ref{firstorder1}) differs in two important respects from the earlier multi-time equation (\ref{biEoM}). First, the Eq.\,(\ref{firstorder1}) is {\it not to be used for noninteracting multipartite systems}, since unphysical correlations would be generated, similarly as pointed out above. This agrees with the expectation, {\it cf.} the discussion in Section\,3., that interacting multipartite systems may be interesting when looking for ontological states and dynamics. 
Second, the Eq.\,(\ref{firstorder1}) is {\it not} obviously discrete time reversal invariant, whereas the second-order Eq.\,(\ref{biEoM}) can be read forwards and backwards alike. However, this can be remedied by applying synchronization ``backwards'', such that 
$\psi_{n_1-1 \; n_2-1}^{\alpha_1\alpha_2}$ features on the right-hand side of 
Eq.\,(\ref{constraint2}) instead.  

\subsection{Specific Hamiltonians (``Ising models'')}   

We finally have to address the question, what kind of interaction Hamiltonian(s) can be chosen in a synchronized first-order equation, such as 
Eq.\,(\ref{firstorder1}), that result in the permutation-like dynamics of $\cal OS$ for a  multipartite system composed of two-state subsystems. 

We consider the two-state subsystems as {\it Ising spins} (or Boolean variables) and propose two $N$-partite working models. The essential requirement here is that the dynamics resulting from the $N$-partite generalization of Eq.\,(\ref{firstorder1}) can be seen as permutations acting on the set of $2^N$ configurations of $N$ Ising spins, the eventual $\cal OS$. 
Thus, superposition states will be avoided.   
\vskip 0.2cm\noindent
{\it Model A}. Employing the Pauli matrix $\hat\sigma_1$, {\it cf.} Eq.\,(\ref{exEoM}), the 
model Hamiltonian is defined by: 
\begin{equation}\label{HA} 
\hat H_A:=
\sum_{i,j=1,\; i<j}^N\; c_{ij}\; 
\hat\sigma_1^{(i)}\otimes\hat\sigma_1^{(j)}\otimes\hat{\mathbf 1}_{(ij)}
\;\;, \end{equation} 
where $\hat{\mathbf 1}_{(ij)}$ is the identity acting on all spins except the pair $(ij)$, 
while the two Pauli matrices flip the state of this pair.  This suitably maps one 
configuration on another one, provided exactly one of the 
coefficients $c_{ij}$  is nonzero in each one of the updating steps. Therefore, the Hamiltonian must  depend on the updating step $n$, $c_{ij}\equiv c_{ij}(n)$. This may represent a situation with a periodically, randomly, or in another way varying {\it external driving} of the dynamics.   
\vskip 0.2cm\noindent 
{\it Model B}. A more interesting situation arises, if the role of the coefficients 
$c_{ij}$ above is played by {\it dynamical link variables} which are Ising spins themselves; we call them {\it edge spins}.  Here we define: 
\begin{equation}\label{HB} 
\hat H_B:=\prod_{i,j=1,\; i<j}^N\;
\big (\textstyle{\frac{1}{2}}[\hat\sigma_3+\hat  1]^{(ij)}\otimes \hat\sigma_1^{(i)}\otimes\hat\sigma_1^{(j)} 
-\textstyle{\frac{1}{2}}[\hat\sigma_3-\hat 1]^{(ij)}\otimes\hat{\mathbf 1}\big )
\;\;, \end{equation}  
with the Pauli matrix $\hat\sigma_3:=\mbox{diag}\{ 1,-1\}$ and where $[\dots ]^{(ij)}$ indicates that the operator in brackets acts on the edge spin   
connecting spins labelled $i$ and $j$. The latter we may call {\it vertex spins} (analogous to matter variables in a lattice gauge theory). The unit operator here refers to vertex spins 
$i$ and $j$. 
Note that the product of terms, instead of a sum, allows to have more than one pair 
of vertex spins flipped, depending on the state of the edge spins, without creating superpositions of spin configurations.  

We remark that 
$(\pm\textstyle{\frac{1}{2}}[\hat\sigma_3\pm\hat 1])^k=\pm\textstyle{\frac{1}{2}}[\hat\sigma_3\pm\hat 1]$, for $k\in\mathbf{N}$ (either $-$ or $+$ signs apply), 
and recall that $\hat\sigma_1^{\;2}=\mathbf{1}$. With the help of these relations and the fact that all operators that constitute $\hat H_B$ commute, one can rewrite this operator in a more familiar looking exponential form (apart from an overall phase): 
\begin{equation}\label{Hexponent} 
-i\hat H_B=\mbox{exp}\Big (-i\textstyle{\frac{\pi}{2}}\sum_{i,j=1,\; i<j}^N\;
\big (\textstyle{\frac{1}{2}}[\hat\sigma_3+\hat  1]^{(ij)}\otimes \hat\sigma_1^{(i)}\otimes\hat\sigma_1^{(j)} 
-\textstyle{\frac{1}{2}}[\hat\sigma_3-\hat 1]^{(ij)}\otimes\hat{\mathbf 1}\big )\Big )
\;. \end{equation} 
Since $\hat\sigma_{1,3}=\hat\sigma_{1,3}^{\;\dagger}$, the Hamiltonian $\hat H_B$ effecting the 
{\it evolution-by-permutation} of the $\cal OS$ (Ising spin configurations) is essentially an  {\it unitary transfer matrix} for a one-step update. 

Of course, {\it Model B} has to be supplemented by a suitable 
dynamics for the edge spins, to be generated by an additional contribution to $\hat H_B$. These edge spins should be updated consistently with the requirements for a  dynamics that qualifies as ontological by permuting $\cal OS$. It leaves much room for model building by specifying the dynamics including the distribution of the links, {\it i.e.} 
which vertex spins are connected {\it via} edge spins and which are not, such as nearest neighbours only, a totally connected, or a random network, {\it etc.}  

It is noteworthy that models of this kind can be made invariant under discrete {\it local Z(2) ``gauge'' transformations}.  
Perhaps this opens the way to build new ontological models with a sort of gauge field dynamics . 

We leave the exploration of the physical relevance of specific models here for future work. However, it may be interesting to point out that certain generalized ``Ising models'' have recently been studied in somewhat related contexts of emergent quantum and gravity or spacetime properties.\,\cite{Wetterich16,Trugenberger15/16} 

\section{Conclusions} 
Embedded in the setting of earlier studies of discrete Hamiltonian Cellular Automata (CA),\,\cite{PRA2014,EmQM13,IJQI16,Discrete14,Wigner13} which we shortly summarize to set the stage, we address here the notion of {\it Ontological States} ($\cal OS$). This has been introduced by 't\,Hooft to form a basis on which ontological models underlying quantum mechanical ones must be built.\,\cite{tHooft2014} 

Characteristic for $\cal OS$, the states a physical system can be in, is that they evolve deterministically by permutations among themselves, since the all too familiar superposition states appearing in quantum mechanics belong to the theory describing experimental findings, but are not considered to exist ``out there''.     

The single-component CA we have studied allowed more or less for the first time to 
reconstruct quantum mechanical models  with nontrivial Hamiltonians in terms of 
deterministic ones that are characterized by a finite discreteness scale. It appears natural to ask, whether in this setting there is room for $\cal OS$ and their particular permutation dynamics. 

We argue that rather simple examples can be constructed in single-component systems, while the extension to interacting multipartite systems composed of two-state subsystems (``Ising spins'') opens new possibilities. 

Earlier we have found that the second-order dynamics of multipartite systems can only be consistently formulated, if a multi-time formulation is 
invoked.\,\cite{IJQI16,Dirac,Tomonaga, Schwinger}. We discuss here in more detail the resulting equations and argue that they tend to produce superposition states that are not ontological. 

In order to nevertheless allow for an evolution-by-permutation of $\cal OS$, we introduce {\it synchronization} constraints for the multi-time dynamics which effectively turn it into a first-order CA updating.  We discuss various Hamiltonians which may lead to physically interesting models, possibly involving a sort of discrete gauge field dynamics, the development of which we leave for future work. -- Thus, we have modified the original framework of the Hamiltonian CA here and ended provisionally with particular forms of Ising models to describe examples of ontological systems. This raises another interesting question which needs to be addressed, in order to complete the scenario of the CA interpretation of  quantum mechanics in the present context. Namely, what would the quantum mechanical description of such ontological systems look like and can it be realized  within the class of CA that we started with? 

\section*{Acknowledgments}
It is a pleasure to thank E. Cohen and G.M. D'Ariano for discussions, C. Wetterich for discussions and kind hospitality at the Institut f\"ur Theoretische Physik (Heidelberg), 
and M. Genovese and P. Aschieri for the invitation to and support during the very enjoyable and stimulating conference {\it Quantum 2017} (Torino) .   


\end{document}